\documentclass[a4paper, 12pt]{article}
\usepackage[toc,page]{appendix}
\usepackage{caption}
\usepackage{makecell}
\usepackage{booktabs}
\usepackage{amsmath}
\usepackage{graphicx}
\usepackage{listings}
\usepackage{color}
\definecolor{dkgreen}{rgb}{0,0.6,0}
\definecolor{gray}{rgb}{0.5,0.5,0.5}
\definecolor{mauve}{rgb}{0.58,0,0.82}
\begin{document}
\title{\textbf{Stock Price Correlation Coefficient Prediction with ARIMA-LSTM Hybrid Model}}
\date{\vspace{0mm}}
\author{
Hyeong Kyu Choi, B.A Student\\
Dept. of Business Administration\\
Korea University\\
Seoul, Korea\\
imhgchoi@korea.ac.kr}
\maketitle
\vspace{10mm}
\title{\centerline{\textbf{Abstract}}} \vspace{3mm} 
Predicting the price correlation of two assets for future time periods is important in portfolio optimization.
We apply LSTM recurrent neural networks (RNN) in predicting the stock price correlation coefficient of two
individual stocks. RNN's are competent in understanding temporal dependencies. The use of LSTM cells
further enhances its long term predictive properties. To encompass both linearity and nonlinearity in the model,
we adopt the ARIMA model as well. The ARIMA model filters linear tendencies in the data and passes on
the residual value to the LSTM model. The ARIMA-LSTM hybrid model is tested against other traditional
predictive financial models such as the full historical model, constant correlation model, single-index model
and the multi-group model. In our empirical study, the predictive ability of the ARIMA-LSTM model turned
out superior to all other financial models by a significant scale. Our work implies that it is worth considering
the ARIMA-LSTM model to forecast correlation coefficient for portfolio optimization. \vspace{2mm} 

\pagenumbering{roman}
\newpage
\tableofcontents
\newpage
\pagenumbering{arabic}
\section{Introduction}\
\indent 
When constructing and selecting a portfolio for investment, evaluation of its expected returns and risks is
considered the bottom line. Markowitz has introduced the Modern Portfolio Theory which proposes methods to
quantify returns and risks of a portfolio, in his paper `Portfolio Selection' (1952) \cite{Mark_Port}. With the derived return and
risk, we draw the efficient frontier, which is a curve that connects all the combination of expected returns and risks
that yield the highest return-risk ratio. Investors then select a portfolio on the efficient frontier, depending on their
risk tolerance.\\
\indent
However, there have been criticisms on Markowitz's assumptions. One of them is that the correlation
coefficient used in measuring risk is constant and fixed. According to Francois Chesnay \& Eric Jondeau's empirical
study on correlation coefficients, stock markets' prices tend to have positive correlations during times of financial
turbulence \cite{Turbulent}. This implies that the correlation of any two assets may as well deviate from mean historical
correlation coefficients subject to financial conditions; thus, the correlation is not stable. Frank Fabozzi, Francis
Gupta and Harry Markowitz himself also briefly discussed the shortcomings of the Modern Portfolio Theory in
their paper, `The Legacy of Modern Portfolio Theory' (2002) \cite{Legacy}.\\

\indent
Acknowledging such pitfalls of the full historical correlation coefficient evaluation measures, numerous
models for correlation coefficient prediction have been devised. One alternative is the Constant Correlation model,
which sets all pairs of assets' correlations equal to the mean of all the correlation coefficients of the assets in the
portfolio \cite{Betas}. Some other forecast models include the Multi-Group model and the Single-Index model. We will
cover these models in our paper at part 2, `Various Financial Models for Correlation Prediction'.
 Although there have been many financial and statistical approaches to estimate future correlation, few have
implemented the neural network to carry out the task. Neural networks are frequently used to predict future stock
returns and have produced noteworthy results\footnote{Y. Yoon, G. Swales (1991) \cite{Yoon_Swales}; A. N. Refenes \textit{et al}. (1994) \cite{Refenes}; K. Kamijo, T. Tanigawa (1990) \cite{Stock_Pattern} ; M. Dixon \textit{et al}. (2017) \cite{Bang}}. Given that stock correlation data can also be represented as time series data -- deriving the correlation coefficient dataset with a rolling time window -- application of neural 
networks in forecasting future correlation coefficients can be expected to have successful results as well. Rather
than circumventing by predicting individual asset returns to compute the correlation coefficient, we cast
predictions directly on the correlation coefficient value itself.\\
\indent
In this paper, we suggest a hybrid model of the ARIMA and the neural network to predict future correlation
coefficients of stock pairs that are randomly selected among the S\&P500 corporations. The model adopts the
Recurrent Neural Network with Long Short-Term Memory cells (for convenience, the model using this cell will
be called LSTM in the rest of our paper). To better predict the time series trend, we also utilize the ARIMA model.
In the first phase, the ARIMA model catches the linear tendencies in the time series data. Then, the LSTM model
attempts to capture nonlinearity in the residual values, which is the output of the former phase. This ARIMA and
neural network hybrid model was discussed in Peter Zhang's literature \cite{Zhang}, and an empirical study was conducted
by James Hansen and Ray Nelson on a variety of time series data \cite{Hansen_Nelson}. The model architecture used in these
literatures are different from what is demonstrated in our paper. We only focus on the hybrid model's versatile
predictive potential to capture both linearity and nonlinearity. Further model details will be elaborated at part 3,
`The ARIMA-LSTM Hybrid Model'.\\
\indent
In the final evaluation step, the ARIMA-LSTM hybrid model will be tested on two time periods which were
not involved in the training step. The layout and methodology of this research will be discussed in detail at part
4, `Research Methodology'. The data will be explored as well in this section. The performance of the model will
then be compared with that of the full historical model as well as other frequently used predictive models that are
introduced in part 2. Finally, the results will be summarized and evaluated in part 5, `Results and Evaluation'.\\
\section{Various Financial Models for Correlation prediction}\
\indent 
 The impreciseness of the full historical model for correlation prediction has largely been acknowledged \cite{Turbulent, Legacy}.
There have been numerous attempts to complement mispredictions. In this section, we discuss three other
frequently used models, along with the full historical model; three of which cited in the literature by Elton \textit{et al}.
(1978) \cite{Betas} -- Full Historical model, Constant Correlation model, and the Single-Index model – and the other, the
Multi-Group model, in another paper of Elton \textit{et al}. (1977) \cite{Multi_Group}. 
\subsection{Full Historical Model}\
\indent
 The Full Historical model is the simplest method to implement for the portfolio correlation estimation. This model adopts the past correlation value to forecast future correlation coefficient. That is, the correlation of two assets for a certain future time span is expected to be equal to the correlation value of a given past period \cite{Betas}.\\
\vspace{3mm}
\begin{center}
\begin{equation}
\hat{\rho}_{ij}^{(t)} = \rho_{ij}^{(t-1)}
\end{equation} 
\end{center}

\indent i, j : asset index in the correlation coefficient matrix \vspace{2mm} \\
However, this model has encountered criticisms on its relative inferior prediction quality compared to other
equivalent models
\vspace{3mm}
\subsection{Constant Correlation Model}\
\indent 
The Constant Correlation model assumes that the full historical model encompasses only the information of
the mean correlation coefficient \cite{Betas}. Any deviation from the mean correlation coefficient is considered a random
noise; it is sufficient to estimate the correlation of each pair of assets to be the average correlation of all pairs of
assets in a given portfolio. Therefore, applying the Constant Correlation model, all assets in a single portfolio have
the same correlation coefficient. \\
\begin{center}
\begin{equation}
\hat{\rho}_{ij}^{(t)} = \frac{\displaystyle \sum_{i>j} \rho_{ij}^{(t-1)}} {\displaystyle n(n-1) / 2}
\end{equation} 
\end{center}

i, j : asset index in the correlation coefficient matrix\\
\indent n : number of assets in the portfolio  \vspace{2mm} \\
\newpage
\subsection{Single-Index Model}\
\indent
The Single-Index model presumes that asset returns move in a systematic way with the `single-index', that is,
the market return \cite{Betas}.
 To quantify the systematic movement with respect to the market return, we need to specify the market return
itself. We call this specification the `market model', which was contrived by H. M. Markowitz \cite{MPT_1950}, and furthered
by Sharpe (1963) \cite{Sharpe}. The `market model' relates the return of asset i with the market return at time t, which is
represented in the following equation:\\
\begin{center}
$R_{i,t} = \alpha_i + \beta_i \hspace{1.5mm} R_{m,t} + \epsilon_{i,t}$
\end{center}

\indent $R_{i,t}$ : return of asset i at time t\\
\indent$R_{m,t}$ : return of the market at time t\\
\indent$\alpha_i$: risk adjusted excess return of asset i \\
\indent$\beta_i$: sensitivity of asset i to the market\\
\indent$\epsilon_{i,t}$: residual return; error term\\
\indent \indent \hspace{2mm} such that, $E(\epsilon_{i}) = 0$ \hspace{.8mm} ; \hspace{.8mm} $Var(\epsilon_{i}) = \sigma^2_{\epsilon_i}$\\

Here, we use the beta($\beta$) of asset i and j to estimate the correlation coefficient. With the equation that,\\
\begin{center}
$Cov( R_i , R_j ) = \rho_{ij} \sigma_i \sigma_j = \beta_i \beta_j \sigma_m^2 $
\end{center} 

\indent $\sigma_i / \sigma_j$ : standard deviation of asset i / j's return \\
\indent $\sigma_m$ : standard deviation of market return\\

The estimated correlation coefficient $\hat{\rho}_{ij}$ would be,
\begin{center}
\begin{equation}
\hat{\rho}_{ij}^{(t)} = \frac{\displaystyle \beta_i \beta_j \sigma_m^2} {\displaystyle \sigma_i \sigma_j}
\end{equation} 
\end{center}
\vspace{3mm}
\subsection{Multi-Group Model}\
\indent
The Multi-Group model \cite{Multi_Group} takes the asset's industry sector into account. Under the assumption that assets in
the same industry sector generally perform similarly, the model sets each correlation coefficient of asset pairs
identical to the mean correlation of the industry sector pair's correlation value. In other words, the Multi-Group
model is a model that applies the Constant Correlation model to each pair of business sectors. For instance, if
company A and company B, each belongs to industry sector $\alpha$ and $\beta$, their correlation coefficient would be the
mean value of all the correlation coefficients of asset pairs with the same industry sector combination ($\alpha, \beta$).\\
\indent
The equation for the prediction is slightly different depending on whether the two industry sectors $\alpha$ and $\beta$
are identical or not. The equation is as follows.
\begin{center}
\begin{equation}
\hat{\rho}_{ij}^{(t)} = \begin{cases}
\frac{\displaystyle \sum_{i \in \alpha}^{n_\alpha} \sum_{j \in \beta ; i \neq j}^{n_\beta} \rho_{ij}^{(t-1)}} {\displaystyle n_\alpha (n_\beta - 1)}, \hspace{2mm} \text{where  $\alpha$ = $\beta$}\\
\frac{\displaystyle \sum_{i \in \alpha}^{n_\alpha} \sum_{j \in \beta ; i \neq j}^{n_\beta} \rho_{ij}^{(t-1)}} {\displaystyle n_\alpha  n_\beta}, \hspace{2mm} \text{where $\alpha$ $\neq$ $\beta$}
\end{cases}
\end{equation}
\end{center}

$\alpha$ / $\beta$ : industry sector notation\\
\indent $n_{\alpha}$ / $n_{\beta}$ : the number of assets in each industry sector 
\vspace{3mm}
\section{The ARIMA-LSTM Hybrid Model}
Time series data is assumed to be composed of the linear portion and the nonlinear portion \cite{Zhang}. Thus, we can
express as follows.\\
\begin{center}
    \begin{equation*}
        x_t = L_t + N_t + \epsilon_t
    \end{equation*}
\end{center}
$L_t$ represents the linearity of data at time $t$, while $N_t$ signifies nonlinearity. The $\epsilon$ value is the error term.\\
\indent 
The Autoregressive Integrated Moving Average (ARIMA) model is one of the traditional statistical models for
time series prediction. The model is known to perform decently on linear problems. On the other hand, the Long
Short-Term Memory (LSTM) model can capture nonlinear trends in the dataset. So, the two models are
consecutively combined to encompass both linear and nonlinear tendencies in the model. The former sector is the
ARIMA model, and the latter is the LSTM model. 
\newpage
\subsection{ARIMA model sector}\
\indent
The ARIMA model is fundamentally a linear regression model accommodated to track linear tendencies in
stationary time series data. The model is expressed as ARIMA(p,d,q). Parameters p, d, and q are integer values
that decide the structure of the time series model; parameter p, q each is the order of the AR model and the MA
model, and parameter d is the level of differencing applied to the data. The mathematical representation of the
ARMA model of order (p,q) is as follows.
\begin{center}
\begin{eqnarray*}
\hat{x}_t & = & c  + \phi_1x_{t-1} + \phi_2x_{t-2} + \cdots + \phi_px_{t-p} \\
          &   & \hspace{2mm}   - \theta_1\varepsilon_{t-1} - \theta_2\varepsilon_{t-2} - \cdots - \theta_q\varepsilon_{t-q}\\
          & = & c + \sum_{k=1}^p \hspace{1mm} \phi_k x_{t-k} - \sum_{l=1}^q \theta_l \hspace{1mm} \varepsilon_{t-l}
\end{eqnarray*}
\end{center}
 Term $c$ is a constant; $\phi_k$ and $\theta_k$ are coefficient values of AR model variable $x_{t-k}$, and MA model variable
$\varepsilon_{t-l}$. $\varepsilon_{t-l}$ is an error notation at period $t–l$ ($\varepsilon_{t-l} = x_{t-l} - \hat{x}_{t-l}$). It is assumed that 
$\varepsilon_{t-l}$ has zero mean with
constant variance, and satisfies the i.i.d condition.\\
\indent 
Box \& Jenkins \cite{Box_Jenkins} introduced a standardized methodology to build an ARIMA model. The methodology
consists of three iterative steps. (1) Model identification and model selection – the type of model, the AR(p) or
MA(q), or ARMA(p,q), is determined. (2) Parameter estimation – the model parameters are adjusted to optimize
the model. (3) Model Checking – residual analysis is performed to better the model.\\
\indent 
In the model identification and model selection step, the proper type of model among the AR and MA models
is decided. To judge which model fits best, stationary time series data needs to be provided. Stationarity requires
that basic statistical properties such as the mean, variance, covariance or autocorrelation be constant over time
periods. In cases of dealing with non-stationary data, differencing is applied once or twice to achieve stationarity
– differencing more than two times is not frequently implemented. After stationarity conditions are satisfied, the
autocorrelation function (ACF) plot and the partial autocorrelation function (PACF) plot are examined to select
the model type.\\
\indent
 The parameter estimation step involves an optimization process utilizing mathematical error metrics such as
the Akaike Information Criterion (AIC), the Bayesian Information Criterion (BIC), or the Hannan-Quinn
Information Criterion (HQIC). In this paper, we resolve to use the AIC metric to estimate parameters.
\begin{center}
$AIC = -2 \hspace{1mm} ln(\hat{L}) + 2k$
\end{center}
\indent The $ln(\hat{L})$ notation is the value of the likelihood function, and $k$ is the degree of freedom, that is, the number of
parameters used. A model that has a small AIC value is generally considered a better model. There are different
ways to compute the likelihood function, $ln(\hat{L})$. We use the maximum likelihood estimator for the computation. This
method tends to be slow, but produces accurate results.
 Lastly, in the model checking step, residual analysis is carried out to finalize the ARIMA model. If residual
analysis concludes that the residual value does not suffice standards, the three steps are iterated until an optimal
ARIMA model is attained. Here, we use the residual that is calculated from the ARIMA model as the input for the 
subsequent LSTM model. As the ARIMA model has identified the linear trend, the residual is assumed to
encompass the non-linear features \cite{Zhang}.\\
\begin{center}
    $x_t - L_t \hspace{2mm} = \hspace{2mm} N_t + \epsilon_t$
\end{center}
The $\epsilon_t$ value would be the final error term of our model.\\
\subsection{LSTM model sector}\
\indent
Neural Networks are known to perform well on nonlinear tasks. Because of its versatility due to large
dimension of parameters, and the use of nonlinear activation functions in each layer, the model can adapt to
nonlinear trends in the data. But empirical studies on financial data show that the performance of neural networks
are rather mixed. For example, in D.M.Q. Nelson \textit{et al}.'s literature \cite{DMQ}, the accuracy of an LSTM neural network
for stock price prediction generally tops other non-neural-network models. However, there are overlapping
portions in the accuracy range of each model, implying that the model not always performs superior to others. This
provides a ground for our paper to use an ARIMA-LSTM hybrid model that encompasses both linearity and
nonlinearity, so as to produce a more sophisticated result compared to pure LSTM neural network models.\\
\indent
 To understand the LSTM model, the mechanism of Recurrent Neural Networks (RNN) should first be
discussed. The RNN is a type of sequential model that performs effectively on time series data. It takes a sequence
of vectors of time series data as input X = [$x_1, x_2, x_3, \cdots, x_t$] and outputs a vector value computed by the
neural network structures in the model's cell, symbolized as A in Figure 1. Vector X is a time series data spanning
t time periods. The values in vector X is sequentially passed through cell A. At each time step, the cell outputs a
value, which is concatenated with the next time step data, and the cell state C. The output value and C serve as
input for the next time step. The process is repeated up to the last time step data. Then, the Backward Propagation
Through Time (BPTT) process, where the weight matrices are updated, initiates. The BPTT process will not be
further illustrated in this paper. For detailed illustration, refer to S. Hochreiter and J. Schmidhuber's literature on
Long Short-Term Memory (1997) \cite{LSTM}.\\
\indent
 The A cell in Figure 1 can be substituted with various types of cells. In this paper, we select the standard LSTM
cell with forget gates, which was introduced by F. Gers \textit{et al}. (1999) \cite{Forget}. The LSTM cell adopted in this paper
comprises four interactive neural networks, each representing the forget gate, input gate, input candidate gate, and
the output gate. The forget gate outputs a vector whose element values are between 0 and 1. It serves as a forgetter
that is multiplied to the cell state $C_{t-1}$ from the former time step to drop values that are not needed and keep those
that are necessary for the prediction.
\begin{center}
    $f_t \hspace{1mm} = \hspace{1mm} \sigma(\hspace{1mm} W_f \cdot [ h_{t-1}, x_t ] + b_f \hspace{1mm})$
\end{center}
The $\sigma$ function, also denoted with the same symbol in Figure 2, is the logistic function, often called the sigmoid.
It serves as the activation function that enables nonlinear capabilities for the model.
\begin{center}
    $\sigma(X) \hspace{1mm} = \hspace{1mm} \frac{\displaystyle 1}{\displaystyle 1 + e^{-X}}$
\end{center}\
\indent
 In the next phase, the input gate and the input candidate gate operate together to render the new cell state $C_t$,
which will be passed on to the next time step as the renewed cell state. The input gate uses the sigmoid as the
activation function and the input candidate utilizes the hyperbolic tangent, each outputting $i_t$ and $\tilde{C}_t$. The $i_t$ 
selects which feature in $\tilde{C}$ should be reflect ed in to the new cell state $C_t$.
\begin{center}
    $i_t \hspace{1mm} = \hspace{1mm} \sigma(\hspace{1mm} W_i \cdot [ h_{t-1}, x_t ] + b_i \hspace{1mm})$ \\ \vspace{2mm}
    $\tilde{C}_t \hspace{1mm} = \hspace{1mm} tanh(\hspace{1mm} W_C \cdot [ h_{t-1}, x_t ] + b_C \hspace{1mm})$
\end{center}
The $tanh$ function, denoted `$tanh$' in Figure 2 as well, is the hyperbolic tangent. Unlike the sigmoid, which
renders value between 0 and 1, the hyperbolic tangent outputs value between -1 and 1.
\begin{center}
    $tanh(X) = \frac{\displaystyle e^{X} - e^{-X}}{\displaystyle e^{X} + e^{-X}}$
\end{center}
\vspace{13pt}
\begin{figure}[h]
\centering
\includegraphics[width=120mm]{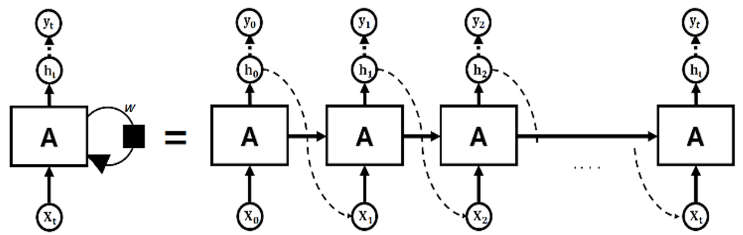}
{\caption*{\textbf{Figure 1.} Structure of Recurrent Neural Network}}
\vspace{20mm}
\includegraphics[width=100mm]{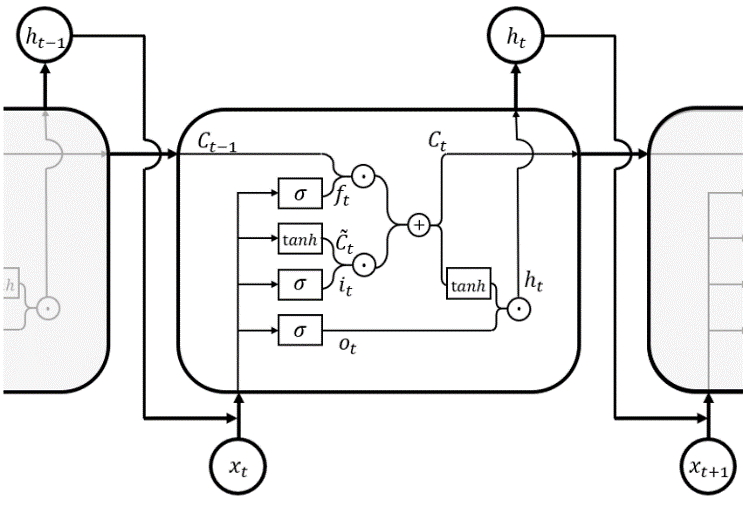}
{\caption*{\textbf{Figure 2.} Inner structure of a Long Short-Term Memory cell}}
\end{figure}\
\indent
 Finally, the output gate decides what values are to be selected, combining $o_t$
 with the $tanh$-applied state $C_t$ as output $h_t$. The new cell state is a combination of the forget-gate-applied former cell state $C_{t-1}$ and the new $tanh$-applied state $C_t$.
\begin{center}
    $o_t \hspace{1mm} = \hspace{1mm} \sigma(\hspace{1mm} W_o \cdot [ h_{t-1}, x_t ] + b_o \hspace{1mm})$ \\ \vspace{2mm}
    $C_t \hspace{1mm} = \hspace{1mm} f_t \hspace{1mm} \cdot \hspace{1mm} C_{t-1} \hspace{2mm} + \hspace{2mm} i_t \hspace{1mm} \cdot \hspace{1mm} \tilde{C}_t$ \\ \vspace{2mm}
    $h_t \hspace{1mm} = \hspace{1mm} o_t \hspace{2mm} \cdot \hspace{2mm} tanh(C_t)$
\end{center}
\indent
 The cell state $C_t$ and output $h_t$ will be passed to the next time step, and will go through a same process.
Depending on the task, further activation functions such as the Softmax or Hyperbolic tangent can be applied to $h_t$'s.
In our paper's case, which is a regression task that has output with values bounded between -1 and 1, we
apply the hyperbolic tangent function to the output of the last element of data vector X. Figure 2 provides a visual
illustration to aid understanding of the LSTM cell inner structure. 
\section{Research Methodology}
\subsection{ARIMA}
\subsubsection{\textit{the Data}}\
\indent
In this paper, we resolve to utilize the ‘adjusted close' price of the S\&P500 firms \footnote{https://en.wikipedia.org/wiki/List\_of\_S\%26P\_500\_companies(accessed 23 May, 2018)}. The price data from 2008-01-01 to 2017-12-31 of the S\&P500 firms are downloaded\footnote{We utilize the ‘Quandl’ API to download stock price data\\ \indent\hspace{3mm}(https://github.com/quandl/quandl-python)}. The data has a small ratio of missing values. The ratio
of missing price data of each asset is around 0.1\%, except for one asset with ticker `MMM', which has a ratio
around 1.1\%. Although MMM's ratio is not that high, missing data imputation seems improbable because the
missing values are found in consecutive days, creating great chasms in the time series. This may cause distortion
when computing the correlation coefficient. So we exclude MMM from our research. For other assets, we impute
the missing data at time t with the value of time t-1 for all assets. Then, we randomly select 150 stocks from the
fully imputed price dataset. The randomly selected 150 firms' tickers are enlisted in `Appendices A'.\\
\indent
 Using the fully imputed 150 set of price data, we compute the correlation coefficient of each pair of assets with
a 100-day time window. In order to add diversity, we set five different starting values, $1^{st}, 21^{st}, 41^{st}, 61^{st}$ and $81^{st}$,
and each apply a rolling 100-day window with a 100-day stride until the end of the dataset. This process renders
55875 sets of time series data ($ _{150}\textrm{C}_2 \cdot 5$), each with 24 time steps. Finally, we generate the train, development,
and test1\&2 data set with the 55875 $\times$ 24 data. We split the data as follows by means to implement the walk-forward
optimization \cite{ladyzynski} in the model evaluation phase.\\

\vspace{1mm}
\begin{tabular}{cll}
\indent $\cdot$ & Train set : & index 1 \hspace{2mm} \textasciitilde \hspace{2mm} 21\\
\indent $\cdot$ & Development set : & index 2 \hspace{2mm} \textasciitilde \hspace{2mm} 22\\
\indent $\cdot$ & Test1 set : & index 3 \hspace{2mm} \textasciitilde \hspace{2mm} 23\\
\indent $\cdot$ & Test2 set : & index 4 \hspace{2mm} \textasciitilde \hspace{2mm} 24\\
\end{tabular}
\begin{figure}[h]
\centering
\includegraphics[width=140mm]{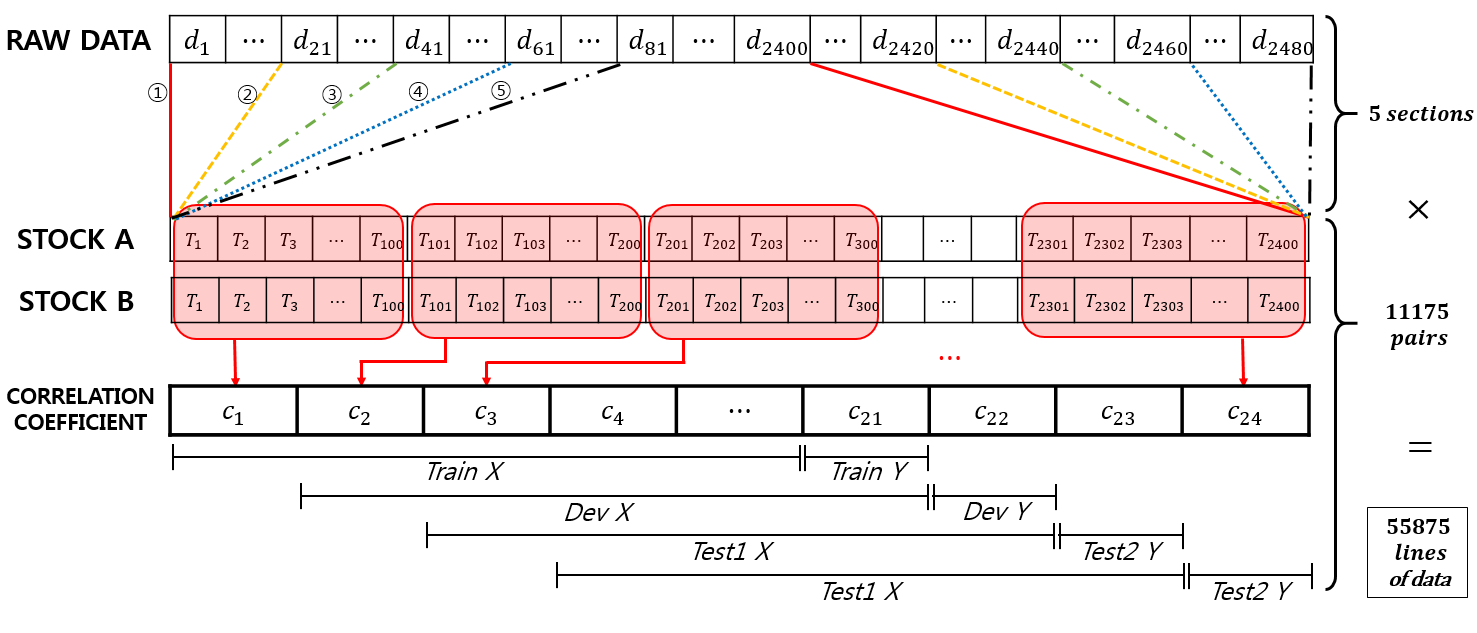}
{\caption*{\textbf{Figure 3.} Data generation scheme}}
\end{figure}
\vspace{15pt}
\subsubsection{\textit{Model Fitting}}\
\indent
Before fitting an ARIMA model, the order of the model must be specified. The ACF plot and the PACF plot
aids the decision process. Most of the datasets showed an oscillatory trend that seemed close to a white noise as
shown in Table 1. Other notable trends includes an increasing/decreasing trend, occasional big dips while steady
correlation coefficient, and having mixed oscillatory-steady periods. Although the ACF/PACF plots indicate that
a great portion of the datasets are close to a white noise, several orders (p, d, q) = (1, 1, 0), (0, 1, 1), (1, 1, 1), (2,
1, 1), (2, 1, 0) seems applicable. We fit the ARIMA model\footnote{We utilize the ‘pyramid’ module to fit ARIMA models\\ \indent \hspace{2mm} (https://github.com/tgsmith61591/pyramid)} with these five orders and select the model with the
least AIC value, for each train/development/test1/test2 dataset's data. The method we use to compute the log
likelihood function for the AIC metric is the maximum likelihood estimator.\\
\indent
 After fitting the ARIMA model, we generate predictions for each 21 time steps to compute the residual value.
Then, the last data point of each data will serve as the target variable Y, and the rest as variable X (Figure 3). The
newly X/Y-split datasets will be the input values for the next LSTM model sector.
\newpage
\begin{figure}[h]
\centering
\includegraphics[width=140mm]{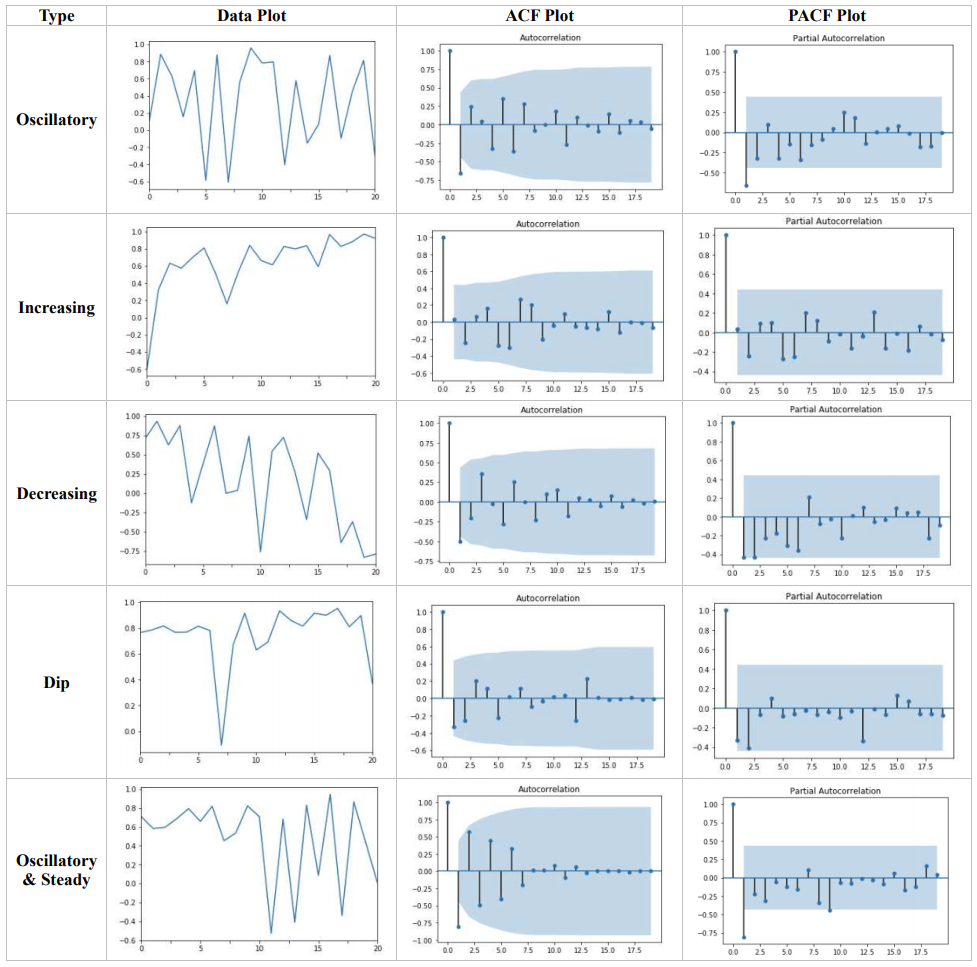}
{\caption*{\textbf{Table 1.} Notable trends and the ACF/PACF of 1-level-differenced data}}
\end{figure}
\newpage
\subsubsection{\textit{the Algorithm}}
\vspace{3mm}
\begin{center}
    \begin{tabular}{l}
    \hline
    \textbf{Algorithm 1.} ARIMA model fitting algorithm\\
    \hline
    \begin{tabular}{rl}
    1: & \textit{datasets} = [train, dev, test1, test2] \\
    2: & \textit{orders} = [ (1,1,0), (0,1,1), (1,1,1), (2,1,1), (2,1,0) ]\\
    3: & \textbf{for all} data \textbf{in} \textit{datasets} \textbf{do}\\
    4: & \hspace{5mm} \textit{X} = empty list\\
    5: & \hspace{5mm} \textit{Y} = empty list\\
    6: & \hspace{5mm} \textbf{for all} time-series \textit{T} \textbf{in} data \textbf{do}\\
    7: & \hspace{5mm}\hspace{5mm} \textit{models} = empty list\\
    8: & \hspace{5mm}\hspace{5mm} \textbf{for all} order \textbf{in} \textit{orders} \textbf{do}\\
    9: & \hspace{5mm}\hspace{5mm}\hspace{5mm} $\textit{M}_{order} = $ fit ARIMA(\textit{T}, order)\\
    10:& \hspace{5mm}\hspace{5mm}\hspace{5mm} \textbf{add}  $\textit{M}_{order}$ \textbf{to} \textit{models}\\
    11:& \hspace{5mm}\hspace{5mm} \textbf{with} least-AIC model $\textit{M}_{\textit{fit}}$ \textbf{in} \textit{models}\\
    12:& \hspace{5mm}\hspace{5mm}\hspace{5mm} \textit{residual} = \textit{X} - predict(T, $\textit{M}_{\textit{fit}}$)\\
    13:& \hspace{5mm}\hspace{5mm} \textbf{add} \textit{residual}[0:20] \textbf{to} \textit{X}\\
    14:& \hspace{5mm}\hspace{5mm} \textbf{add} \textit{residual}[20] \textbf{to} \textit{Y}\\
    15:& \hspace{5mm} \textbf{save} \textit{X, Y}\\
    \hline
    \end{tabular}
    \end{tabular}
\end{center}
\vspace{3mm}
\subsection{LSTM}
\subsubsection{\textit{the Data}}\
\indent
We use the residual values, derived from the ARIMA model, of the 150 randomly selected S\&P500 stocks as
input for the LSTM model. The datasets include the train X/Y, development X/Y, test1 X/Y, and test2 X/Y. Each
X dataset has 55875 lines with 20 time steps, with a corresponding Y dataset for each time series (Figure 3). The data points are generally around 0, as the input is a residual dataset (Figure 4). 
\begin{figure}
\centering
\includegraphics[width = 80mm]{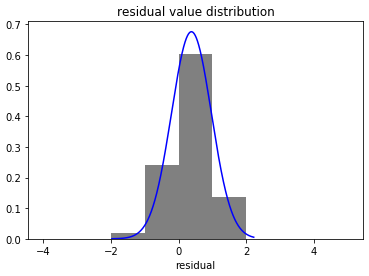}
\caption*{\textbf{Figure 4.} Data Point Distribution}
\end{figure}
\subsubsection{\textit{Model Training}} \
\indent
The architecture of the model for our task is an RNN neural network that employs 25 LSTM units\footnote{We utilize the ‘keras’ module to train the LSTM model\\ \indent \hspace{3mm} (https://github.com/keras-team/keras)}. The last outputs of 25 LSTM units is merged into a single value with a fully connected layer. Then, the value will be passed through a doubled-hyperbolic tangent activation function to output a single final prediction. The doubled-hyperbolic tangent is simply the hyperbolic tangent function scaled by a factor of 2. Figure 5 shows a simplified
architecture of the model.\\
\indent
 When training the model, it is crucial to keep an eye on overfitting. Overfitting occurs when the model fits
excessively on dataset while training. Hence, the predictive performance on the train dataset will be high, but will
be poor on other newly introduced data. To monitor this problem, a separate set of development dataset is used.
We train the LSTM model with the train dataset until the predictive performances on the train dataset and
development dataset become similar to each other.\\
\indent
 The dropout method is one of the widely used methods to prevent overfitting. It prevents the neurons to develop
interdependency, which causes overfitting. This is executed by simply turning off neurons in the network during
training with probability $p$. Then, in the testing phase, dropout is disabled and each weight values are multiplied
by $p$, to scale down the output value into a desired boundary. Moreover, dropout has the effect of training multiple
neural networks and averaging the outputs \cite{Dropout}.\\
\indent
 Other than dropout, we considered more regularization to prevent overfitting. There are mainly two types of
regularization methods: Lasso regularization (L1) and Ridge regularization (L2). These regularizers keep the
weight values of each network in the LSTM model from becoming too large. Big parameter values of each layers
may cause the network to focus severely on few features, which may result in overfitting. 
A general expression of the error function with regularization is as follows.\\
\begin{center}
\begin{equation*}
    \sum_{i=1}^{n} \{Y_i - (W \cdot X_i +b)\}^2 + \lambda_W\sum_{i=1}^k \sum_{j=1}^{k'} W_{ij}^2 
     + \lambda_b \sum_{i=1}^l \sum_{j=1}^{l'} b_{ij}^2
\end{equation*}
\end{center}
\vspace{3mm}
Parameters $\lambda_{\textit{W}}$ and $\lambda_{\textit{b}}$ determine the intensity of regularization of the cost function. If the lambda values are too high, the model will be under-trained. On the other hand, if they are too low,
regularization affect will be minimal. 
In our model, after trial and error, it turned out that not applying any regularization performs better. We tried more complex architectures with regularization, but for all architectures, models with no regularization had superior outputs.\\
\indent
 Another problem to pay attention to when training a neural network model is the vanishing/exploding gradient.
This is particularly emphatic for RNN's. Due to a deep propagation through time, the gradients far away from the
output layer tend to be very small or large, preventing the model from training properly. The remedy for this
problem is the LSTM cell itself. The LSTM is capable of connecting large time intervals without loss of
information \cite{LSTM}.\\
\indent
 Other miscellaneous details about the training process includes the use of mini-batch of size 500, the ADAM
optimization function \textit{et cetera}. For detail, refer to the LSTM section source codes in `Appendices B'.
\begin{figure}[h]
\vspace{15mm}
\centering
\includegraphics[width=140mm]{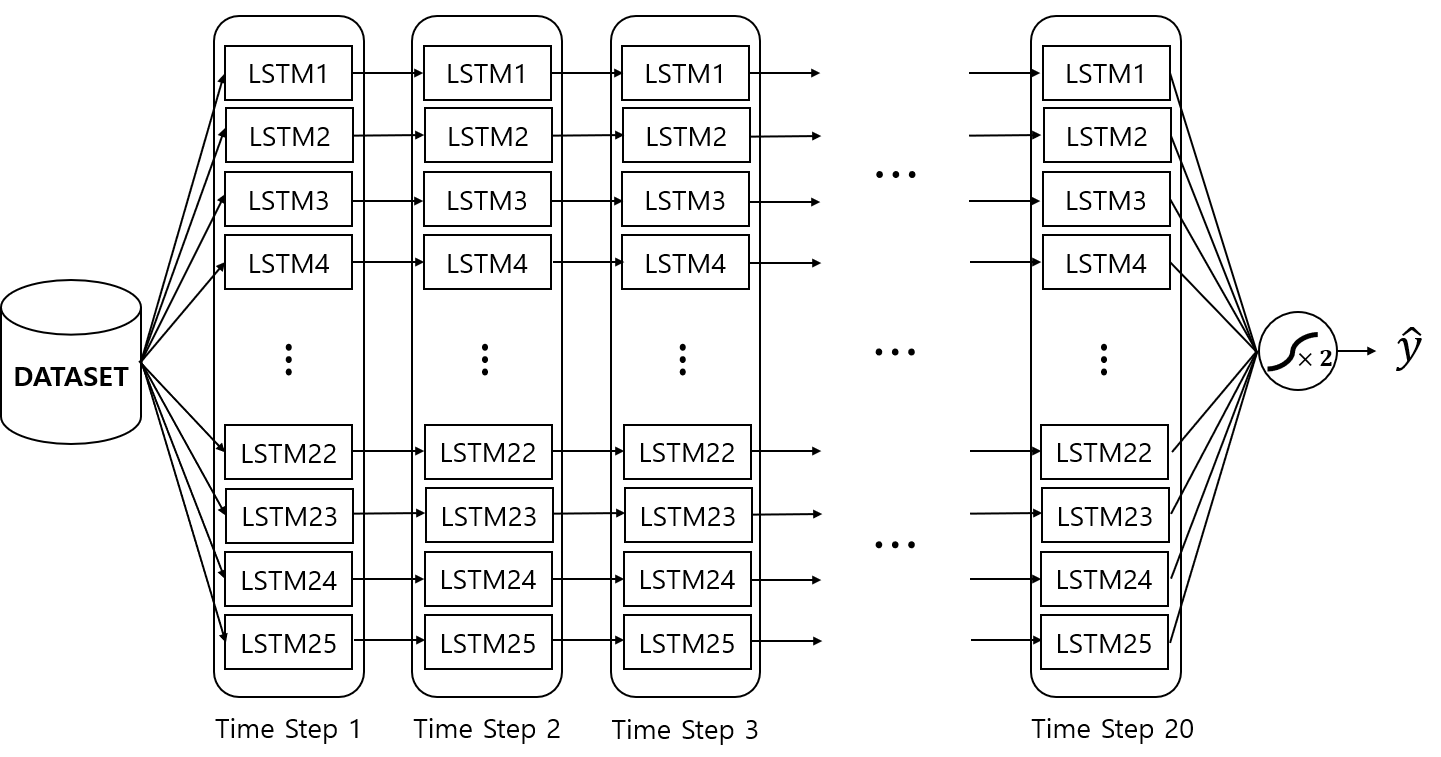}
{\caption*{\textbf{Figure 5.} LSTM model sector architecture}}
\end{figure}
\subsubsection{\textit{Evaluation}}\
\indent
The walk-forward optimization method \cite{ladyzynski} is used as the evaluation method. The walk-forward optimization
requires that a model be fitted for each rolling time intervals. Then, for each time interval, the newly trained model
is tested on the next time step. This ensures the robustness of the model fitting strategy. However, this process is
computationally expensive. In addition, our paper's motive is to fit parameters of a model that generalizes well on
various assets as well as on different time periods. Thus, it is needless to train multiple models to approve of the
model-fitting strategy. Rather than training a new model for each rolling train-set window, we resolve to train a
single model with the first window and apply it to three time intervals – the development set and the test1/test2
set.\\
\indent
 We selected our optimal model with the Mean Squared Error (MSE) metric. That is, the cost function of our
model was the MSE. For further evaluation, the Mean Absolute Error (MAE) and Root Mean Squared Error(RMSE)
was also investigated.
\begin{center}
        MSE $= \frac{\displaystyle 1}{\displaystyle n}$  $\sum_{i=1}^n (y_i - \hat{y}_i)^2$\\ \vspace{2mm}
        MAE $= \frac{\displaystyle 1}{\displaystyle n}$  $\sum_{i=1}^n |y_i - \hat{y}_i|$\\
\end{center}
The selected optimal model is then tested on two recent time periods. We use two separate datasets to test the
model because the development set is deemed to be involved in the learning process as well.\\
\indent
 If the model's correlation coefficient prediction on two time periods turn out decent as well, we then test our
model against former financial predictive models. The MSE and MAE values are computed for the four financial
models as well. For the constant correlation model and the multi-group model, we regarded the 150 assets we
selected randomly to be our portfolio constituents.
\subsubsection{\textit{the Algorithm}}
\vspace{3mm}
\begin{center}
    \begin{tabular}{l}
    \hline
    \textbf{Algorithm 2.} LSTM model training algorithm\\
    \hline
    \begin{tabular}{rl}
    1: & \textbf{read} [train\_X/Y, dev\_X/Y, test1\_X/Y, test2\_X/Y] \\
    2: & \textbf{define} \textit{model}\\
    3: & \hspace{5mm} \textbf{add} LSTM(units = 25)\\
    4: & \hspace{5mm} \textbf{add} Dense(shape=(25,1), activation=`double-tanh')\\
    5: & \textbf{Repeat}\\
    6: & \hspace{5mm} \textbf{Forward\_propagate} \textit{model} \textbf{with} train\_X\\
    7: & \hspace{5mm} \textbf{Backward\_propagate} \textit{model} \textbf{with} train\_Y\\
    8: & \hspace{5mm} \textbf{Update} \textit{model} parameters\\
    9: & \hspace{5mm} train\_MSE, train\_MAE = \textit{model}(train\_X, train\_Y)\\
    10:& \hspace{5mm} dev\_MSE, dev\_MAE = \textit{model}(dev\_X, dev\_Y) \\
    11:& \hspace{5mm} \textbf{if}  train\_MSE, dev\_MSE converged \\
    12:& \hspace{5mm}\hspace{5mm} \textbf{end Repeat}\\
    13:& test1\_MSE, test1\_MAE = \textit{model}(test1\_X, test1\_Y) \\
    14:& test2\_MSE, test2\_MAE = \textit{model}(test2\_X, test2\_Y) \\
    \hline
    \end{tabular}
    \end{tabular}
\end{center}
\vspace{3mm}
\section{Results And Evaluation}\
\indent
After around 200 epochs, the train dataset's MSE value and development dataset's MSE value started to
converge (Figure 6). The MAE learning curve showed a similar trend as well. Among the models, we selected the
$247^{th}$ epoch's model. The epoch was decided based on both the overfitting metric and the performance metric. The overfitting metric was represented with the normalized value of the MSE difference between the train \& development dataset. And the performance metric was represented with the normalized value of the MSE sum of the train \& development datset. Then, the sum of the two normalized value was calculated to find the epoch that had the least value. The mathematical representation of the criterion is as follows.\\
\begin{center}
    $criterion = \frac{\displaystyle diff_{MSE} - mean(diff_{MSE})}{\displaystyle stdev(diff_{MSE})} + 
    \frac{\displaystyle sum_{MSE} - mean(sum_{MSE})}{\displaystyle stdev(sum_{MSE})}$
\end{center} \

\indent
 With the selected ARIMA-LSTM hybrid model, the MSE, RMSE and MAE values of the prediction were
calculated. The MSE value on the development, test1, and test2 dataset were 0.1786, 0.1889, 0.2154 each. The
values have small variations, which means the model has been generalized adequately.\\
\indent
 Then, the metric values were compared with that of other financial models. Among the financial models, the Constant Correlation model performed the best on our 150 S\&P500 stocks' dataset, just as what the empirical
study of E. J. Elton \textit{et al}. has shown \cite{Betas}. However, its performance was nowhere near the ARIMA-LSTM hybrid
model's predictive capacity. The ARIMA-LSTM's MSE value was nearly two thirds of that of other equivalent models. The MAE metric also showed clear outperformance. Table 2
demonstrates all metrics' values for every dataset, for each model. The least value of each metric was boldfaced.
Here, we can easily notice the how all the metric values of the ARIMA-LSTM model are in boldface.\\
\indent
 For further investigation, we tested our final model on different assets in the S\&P500 firms. Excluding the 150
assets we have already selected to train our model, we randomly selected 10 assets and generated datasets with
identical structures as the ones used in the model training and testing. This generates 180 lines of data. We then
pass the data into our ARIMA-LSTM hybrid model and evaluate the predictions with the MSE, RMSE and MAE
metrics. We iterate this process 10 times to check for model stability. The output of 10 iterations are demonstrated
in Table 3.\\
\indent
 The MSE values of 10 iterations range from 0.1447 to 0.2353. Although there is some variation in the results
compared to the Test1 \& 2, this may be due to a relatively small sample size, and the outstanding performance of
the model makes it negligible. Therefore, we may carefully affirm that our ARIMA-LSTM model is robust.
\begin{figure}[h]
\centering
\includegraphics[width=140mm]{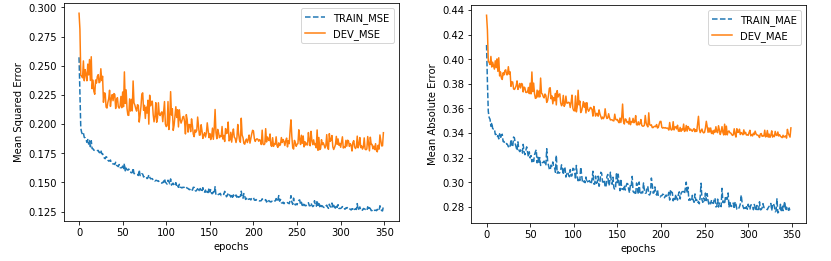}
{\caption*{\textbf{Figure 6.} Learning curves of the ARIMA-LSTM model training process }}
\end{figure}
\begin{center}
\begin{table}[htbp]
\makebox[\textwidth][c]{
    \begin{tabular}{c|ccc|ccc|ccc}
\toprule
    & \multicolumn{3}{c|}{Development dataset} & \multicolumn{3}{c|}{Test1 dataset} & \multicolumn{3}{c}{Test2 dataset} \\
\midrule
    & MSE & RMSE & MAE & MSE & RMSE & MAE & MSE & RMSE & MAE \\ \hline
\textbf{ARIMA-LSTM} & \textbf{.1786} & \textbf{.4226} & \textbf{.3420} & \textbf{.1889} & \textbf{.4346} & \textbf{.3502} & \textbf{.2154} & \textbf{.4641} & \textbf{.3735}\\
Full Historical & .4597& .6780 &.5449 &.5005 &.7075 &.5741& .4458 &.6677& .5345 \\
Constant Correlation & .2954& .5435& .4423 &.2639& .5137 &.4436 &.2903 &.5388 &.4576 \\
Single-Index &.4035 &.6352 &.5165& .3517 &.5930 &.4920& .3860 &.6213 &.5009\\
Multi-Group &.3079 &.5549& .4515 &.2910 &.5394 &.4555 &.2874& .5361 &.4480 \\
\bottomrule
\end{tabular}
}
\caption*{\textbf{Table 2.} ARIMA-LSTM model performance results and its comparison}
\end{table}
\end{center}
\begin{center}
\begin{table}[htbp]
\makebox[\textwidth][c]{
    \begin{tabular}{c|c|c|c|c}
\toprule
Iter.   & Tickers & MSE & RMSE & MAE \\
\midrule
1 & \makecell{PRGO, MRO, ADP, HCP, FITB,\\PEG, SYMC, EOG, MDT, NI} & .2025 &.4500 &.3732 \\ \hline
2 & \makecell{STI, COP, MCD, AON, JBHT,\\DISH, GS, LRCX, CTXS, LEG} & .1517 &.3895 &.3331 \\ \hline
3 & \makecell{TJX, EMN, JCI, C, BIIB,\\HOG, PX, PH, XEC, JEC} & .1680 &.4099 &.3476 \\ \hline
4 & \makecell{ROP, AZO, URI, TROW, CMCSA,\\SLB, VZ, MAC, ADS, MCK} & .1966  &.4434 &.3605 \\ \hline
5 & \makecell{RL, CVX, SRE, PFE, PCG,\\UTX, NTRS, INCY, COP, HRL} & .2353  &.4851 &.3951 \\ \hline
6 & \makecell{FE, STI, EA, AAL, XOM,\\JNJ, COL, APC, MCD, VFC} & .2175  &.4664 & .3709 \\ \hline
7 & \makecell{BBY, AXP, CAG, TGT, EMR,\\MNST, HSY, MCK, INCY, WBA} & .1447  & .3804 & .3094 \\ \hline 
8 & \makecell{BXP, HST, NI, ESS, GILD,\\TSN, T, MSFT, LEG, COST} & .1997  &.4469 &.3518 \\ \hline
9 & \makecell{CVX, FE, WMT, IDXX, GOOGL,\\PKI, EQIX, DISH, FTI, HST} & .1785  &.4225 &.3331 \\ \hline
10 & \makecell{NKE, VAR, DVN, VRSN, PFG,\\HAS, UNP, EQT, FE, AIV} & .2168  &.4656 & .3742  \\
\bottomrule
\end{tabular}
}
\caption*{\textbf{Table 3.}  ARIMA-LSTM testing results on different asset combinations }
\end{table}
\end{center}
\section{Conclusion}\
\indent
The purpose of our empirical study was to propose a model that performs superior to extant financial
correlation coefficient predictive models. We adopted the ARIMA-LSTM hybrid model in an attempt to first filter
out linearity in the ARIMA modeling step, then predict nonlinear tendencies in the LSTM recurrent neural network.
The testing results showed that the ARIMA-LSTM hybrid model performs far superior to other equivalent
financial models. Model performance was validated on both different time periods and on different combinations
of assets with various metrics such as the MSE, RMSE, and the MAE. The values nearly halved that of the Constant
Correlation model, which, in our experiment, turned out to perform best among the four financial models. Judging
from such outperformance, we may presume that the ARIMA-LSTM hybrid model has sufficient predictive
potential. Thus, the ARIMA-LSTM model as a correlation coefficient predictor for portfolio optimization would
be considerable. With a better predictor, the portfolio is optimized more precisely, thereby enhancing returns in
investments.\\
\indent
 However, our experiment did not cover time periods before the year 2008. So our model may be susceptible
to specific financial conditions that were not present in the years between 2008 and 2017. But financial anomalies
and noises are always prevalent. It is impossible to embrace all probable specific tendencies into the model. Hence,
further research into dealing with financial black swans is called for. 
\newpage
\begin{appendices}
\section{150 S\&P500 Stocks}
* This is the list of tickers of the 150 randomly selected S\&P500 stocks. 
\begin{center}
\vspace{5mm}
\begin{tabular}{|c|c|c|c|c|c|}
\hline
CELG &PXD &WAT& LH& AMGN& AOS\\ \hline
EFX &CRM &NEM& JNPR& LB& CTAS\\ \hline
MAT& MDLZ &VLO &APH &ADM& MLM\\ \hline
BK &NOV& BDX &RRC& IVZ &ED\\ \hline
SBUX &GRMN&CI &ZION &COO &TIF\\ \hline
RHT& FDX& LLL &GLW &GPN &IPGP\\ \hline
GPC& HPQ& ADI& AMG &MTB& YUM\\ \hline
SYK &KMX &AME& AAP& DAL& A\\ \hline
MON &BRK &BMY &KMB &JPM &CCI\\ \hline
AET &DLTR& MGM& FL& HD& CLX\\ \hline
OKE &UPS &WMB &IFF &CMS &ARNC\\ \hline
VIAB &MMC& REG &ES &ITW &NDAQ\\ \hline
AIZ &VRTX &CTL& QCOM& MSI &NKTR\\ \hline
AMAT &BWA &ESRX &TXT &EXR &VNO\\ \hline
BBT &WDC &UAL &PVH &NOC &PCAR\\ \hline
NSC& UAA& FFIV &PHM &LUV &HUM\\ \hline
SPG &SJM& ABT &CMG& ALK& ULTA\\ \hline
TMK &TAP &SCG &CAT &TMO&AES\\ \hline
MRK &RMD& MKC& WU &CAN &HIG\\ \hline
TEL& DE& ATVI& O &UNM &VMC\\ \hline
ETFC &CMA &NRG& RHI& RE& FMC\\ \hline
MU &CB& LNT& GE &CBS& ALGN\\ \hline
SNA &LLY &LEN &MAA &OMC&F\\ \hline
APA &CDNS& SLG& HP &XLNX& SHW\\ \hline
AFL &STT &PAYX &AIG& FOX& MA \\ \hline
\end{tabular}
\end{center}
\newpage
\section{LSTM Model Source Code}
* This source code is a simplified version; unnecessary portions were contracted or omitted. For original and other
relevant source codes, visit \\`https://github.com/imhgchoi/Corr\_Prediction\_ARIMA\_LSTM\_Hybrid'. 
\vspace{30pt}
\begin{lstlisting}
import pandas as pd
import numpy as np
import os
from keras.models import Sequential, load_model
from keras.layers import Dense, LSTM, Activation
from keras import backend as K
from keras.utils.generic_utils import get_custom_objects
from keras.callbacks import ModelCheckpoint
from keras.regularizers import l1_l2


# Train - Dev - Test Generation
train_X= pd.read_csv('~/train_dev_test/after_arima/train_X.csv')
dev_X = pd.read_csv('~/train_dev_test/after_arima/dev_X.csv')
test1_X = pd.read_csv('~/train_dev_test/after_arima/test1_X.csv')
test2_X = pd.read_csv('~/train_dev_test/after_arima/test2_X.csv')
train_Y = pd.read_csv('~/train_dev_test/after_arima/train_Y.csv')
dev_Y = pd.read_csv('~/train_dev_test/after_arima/dev_Y.csv')
test1_Y = pd.read_csv('~/train_dev_test/after_arima/test1_Y.csv')
test2_Y = pd.read_csv('~/train_dev_test/after_arima/test2_Y.csv')


# data sampling
STEP = 20
_train_X = np.asarray(train_X).reshape((int(1117500/STEP), 20, 1))
_dev_X = np.asarray(dev_X).reshape((int(1117500/STEP), 20, 1))
_test1_X = np.asarray(test1_X).reshape((int(1117500/STEP), 20, 1))
_test2_X = np.asarray(test2_X).reshape((int(1117500/STEP), 20, 1))
_train_Y = np.asarray(train_Y).reshape(int(1117500/STEP), 1)
_dev_Y = np.asarray(dev_Y).reshape(int(1117500/STEP), 1)
_test1_Y = np.asarray(test1_Y).reshape(int(1117500/STEP), 1)
_test2_Y = np.asarray(test2_Y).reshape(int(1117500/STEP), 1)



#define custom activation
class Double_Tanh(Activation):
    def __init__(self, activation, **kwargs):
    super(Double_Tanh, self).__init__(activation, **kwargs)
    self.__name__ = 'double_tanh'

def double_tanh(x):
    return (K.tanh(x) * 2)
get_custom_objects().update({'double_tanh':Double_Tanh(double_tanh)})


# Model Generation
model = Sequential()
model.add(LSTM(25, input_shape=(20,1)))
model.add(Dense(1))
model.add(Activation(double_tanh))
model.compile(loss='mean_squared_error', optimizer='adam', metrics=['mse', 'mae']) 

# Fitting the Model
model_scores = {}
epoch_num=1
for _ in range(50):
    # train the model
    dir = '~/models/hybrid_LSTM'
    file_list = os.listdir(dir)
    if len(file_list) != 0 :
        epoch_num = len(file_list) + 1
        recent_model_name = 'epoch'+str(epoch_num-1)+'.h5'
        filepath = '~/models/hybrid_LSTM/'+recent_model_name
        model = load_model(filepath)
    filepath = '~/models/hybrid_LSTM/epoch'+str(epoch_num)+'.h5'
    checkpoint = ModelCheckpoint(filepath, monitor='loss', verbose=1, save_best_only=False, mode='min')
    callbacks_list = [checkpoint]
    if len(callbacks_list) == 0:
        model.fit(_train_X, _train_Y, epochs=1, batch_size=500, shuffle=True)
    else:
        model.fit(_train_X, _train_Y, epochs=1, batch_size=500, shuffle=True, callbacks=callbacks_list)
        
        
# test the model
score_train = model.evaluate(_train_X, _train_Y)
score_dev = model.evaluate(_dev_X, _dev_Y)

# get former score data
df = pd.read_csv('~/models/hybrid_LSTM.csv')
train_mse = list(df['TRAIN_MSE'])
dev_mse = list(df['DEV_MSE'])
train_mae = list(df['TRAIN_MAE'])
dev_mae = list(df['DEV_MAE'])

# append new data
train_mse.append(score_train[1])
dev_mse.append(score_dev[1])
train_mae.append(score_train[2])
dev_mae.append(score_dev[2])

# organize newly created score dataset
model_scores['TRAIN_MSE'] = train_mse
model_scores['DEV_MSE'] = dev_mse
model_scores['TRAIN_MAE'] = train_mae
model_scores['DEV_MAE'] = dev_mae

# save newly created score dataset
model_scores_df = pd.DataFrame(model_scores)
model_scores_df.to_csv('~/models/hybrid_LSTM.csv') 
\end{lstlisting}
\end{appendices}
\vspace{1cm}
\section*{Acknowledgement}
We thank developers of the `Quandl' API, `Pyramid-arima' module, and `Keras' module, who provided open source codes that alleviated the burden of our research. \\
We also thank an anonymous commenter with a pseudo-name 'Moosefly', who discovered a crucial error in the ARIMA modeling section source code.
\newpage
\bibliographystyle{plain}

\begin{thebibliography}{10}

\bibitem{Refenes}
G.~Francis A.~N.~Refenes, A.~Zapranis.
\newblock Stock performance modeling using neural networks: A comparative study
  with regression models.
\newblock {\em Neural Networks}, 7(2):375--388, 1994.

\bibitem{DMQ}
R.A. de~Oliveira D.M.Q.~Nelson, A.C.M.~Pereira.
\newblock Stock market’s price movement prediction with lstm neural networks.
\newblock {\em In 2017 International Joint Conference on Neural Networks
  (IJCNN)}, pages 1419--1426, 2017.

\bibitem{Betas}
T.~J.~Urich E.~J.~Elton, M. J.~Gruber.
\newblock Are betas best?
\newblock {\em Journal of Finance}, 33:1375--1384, 1978.

\bibitem{MPT_1950}
M.~J.~Gruber E.J.~Elton.
\newblock Modern portfolio theory, 1950 to date.
\newblock {\em Journal of banking And Finance}, 21:1743--1759, 1997.

\bibitem{Multi_Group}
M.~W.~Padberg E.J.~Elton, M. J.~Gruber.
\newblock Simple rules for optimal portfolio selection: The multi group case.
\newblock {\em Journal of Financial and Quantitative Analysis}, 12(3):329--349,
  1977.

\bibitem{Turbulent}
E.~Jondeau F.~Chesnay.
\newblock Does correlation between stock returns really increase during
  turbulent periods?
\newblock {\em Economic Notes}, Vol.30, no.1-2001:53--80, 2001.

\bibitem{Legacy}
H.~M.~Markowitz F.~J.~Fabozzi, F.~Gupta.
\newblock The legacy of modern portfolio theory.
\newblock {\em The Journal of Investing}, Vol. 11, No. 3:7--22, 2002 Fall.

\bibitem{Forget}
F.~Cummins F.A.~Gers, J.~Schmidhuber.
\newblock Learning to forget: Continual prediction with lstm.
\newblock {\em Technical Report}, IDSIA-01-99, 1999.

\bibitem{Box_Jenkins}
G.~Jenkins G.E.P.~Box.
\newblock Time series analysis, forecasting and control.
\newblock {\em Holden-Day, San Francisco, CA}, 1970.

\bibitem{Hansen_Nelson}
R.~D.~Nelson J.V.~Hansen.
\newblock Time-series analysis with neural networks and arima-neural network
  hybrids.
\newblock {\em Journal of Experimental And Theoretical Artificial
  Intelligence}, 15:3:315--330, 2003.

\bibitem{Stock_Pattern}
T.~Tanigawa K.~Kamijo.
\newblock Stock price pattern recognition - a recurrent neural network
  approach.
\newblock {\em 1990 IJCNN International Joint Conference on Neural Networks},
  vol.1, San Diego, CA, USA:215--221, 1990.

\bibitem{Bang}
J.~H.~Bang M.~Dixon, D.~Klabjan.
\newblock Classification-based financial markets prediction using deep neural
  networks.
\newblock {\em Algorithmic Finance}, 6(3-4):67--77, 2017.

\bibitem{Mark_Port}
H.~M. Markowitz.
\newblock Portfolio selection.
\newblock {\em The Journal of Finance}, Vol.7, No.1:77--91, Mar. 1952.

\bibitem{Dropout}
A.~Krizhvsky I. Sutskever R.~Salakhutdinov N.~Srivastave, G.~Hinton.
\newblock Dropout: A simple way to prevent neural networks from overfitting.
\newblock {\em Journal of Machine Learning Research}, 15:1929--1958, 2014.

\bibitem{ladyzynski}
P.~Grzegorzewski P.~ladyzynski, K.~Zbikowski.
\newblock Stock trading with random forests, trend detection tests and force
  index volume indicators.
\newblock {\em Artificial Intelligence and Soft Computing, Lecture Notes in
  Computer Science}, 7895:pp.441--452, 2013.

\bibitem{LSTM}
J.~Schmidhuber S.~Hochreiter.
\newblock Long short-term memory.
\newblock {\em Neural Computation}, 9(8):1735--1780, 1997.

\bibitem{Sharpe}
W.F. Sharpe.
\newblock A simplified model for portfolio analysis.
\newblock {\em Management Science}, 13:277--293, 1963.

\bibitem{Yoon_Swales}
G.~Swales Y.~Yoon.
\newblock Predicting stock price performance: A neural network approach.
\newblock {\em System Sciences, Proceedings of the Twenty-Fourth Annual Hawaii
  International Conference on}, Vol. 4, IEEE:156--162, 1991.

\bibitem{Zhang}
G.P. Zhang.
\newblock Time series forecasting using a hybrid arima and neural network
  model.
\newblock {\em Neurocomputing}, 50:159--175, 2003.

\bibitem{misc}
F.~Zhao.
\newblock Forecast correlation coefficient matrix of stock returns in portfolio
  analysis, 2013.

\end{thebibliography}

\nocite{misc}
\end{document}